# Analysis of the dynamical evolution of the Quadrantid meteoroid stream

G.E. Sambarov[1,2], T.Yu. Galushina[2], O.M. Syusina[2]

[1]*LSR Group, Kazanskaya st. 36, 190031, St. Petersburg, Russian Federation, e-mail detovelli@gmail.com (G.E. Sambarov)*

[2]*Tomsk State University, Lenina pr, 36, 634050, Tomsk, Russian Federation, phone +73822529776, e-mail volna@sibmail.com (T.Yu.Galushina).*

____________________________________________________________________________________

**Abstract**

We investigate numerically the dynamical evolution of simulated meteoroid stream of the Quadrantids ejected from the parent body of the asteroid (196256) 2003 EH1. The main goal of this work is to identify mean motion and secular resonances and to study the mutual influence of resonance relations and close encounters with the major planets. Since the dynamics of this asteroid is predictable only on short time intervals, and not only close and/or multiple close encounters with major planets, but also the presence of at least one unstable resonance can lead to chaotic motion of test particles, we studied their resonant dynamics. The dynamical evolution of the test particles expects possible scenario for resonant motion. We conjecture that the reasons of chaos are the overlap of stable secular resonances and unstable mean motions resonances and close and/or multiple close encounters with the major planets. The estimate of the stability of orbits in which the particles in simulations moved was carried out by analyzing the behavior of the parameter MEGNO (Mean Exponential Growth factor of Nearby Orbits). The larger part of the identified resonances is stable. We found a peculiar behavior for this stream. Here, we show that the orbits of some ejected particles are strongly affected by the Lidov–Kozai mechanism that protects them from close encounters with Jupiter. Lack of close encounters with Jupiter leads to a rather smooth growth in the parameter MEGNO and the behavior imply the stable motion of simulation particles of the Quadrantids meteoroid stream.

**Key words:** Asteroid; dynamical evolution; (196256) 2003 EH1; the Quadrantids meteor stream; resonances; the Lidov–Kozai mechanism.

____________________________________________________________________________________

## 1. Introduction

Meteoroid stream may have a hazardous impact on a spacecraft. For example, the ESA communications satellite Olympus lost pointing control in 1993 likely due to the impact of the Perseid meteoroids (Caswell, 1995). During the same year, due to the Perseids, NASA was forced to reverse the Hubble Space Telescope direction to avoid the lens damage. That is why it is necessary to study dynamical properties of meteoroid streams. Another likely reason is that there is something about the dynamical phenomena themselves we learn. To model a meteoroid stream formation, a precise parent body orbit is needed, only then it is advisable to analyze its dynamical evolution.

The Quadrantid shower is one of the most intense visual January showers, but it has only become active recently, being first noticed around 1835 AD. This discovery is variously attributed to Wartmann, Quetelet and Herrick (Quetelet, 1839; Fisher, 1930). The name of the shower originates from the Quadrans Muralis constellation. This is now a defunct constellation but it existed when the stream was recognized in 1835 by Quetelet (Fisher, 1930). We can assume that Quadrantid meteors had not been seen any time before then. The mass sorting that occurred in the Quadrantids flow indicates large-scale perturbations (Hughes et.al., 1979). These were quantified by Hughes & Taylor (1977).

The core of the Quadrantid is only 200-300 years old and is associated with asteroid (196256) 2003 EH1 (Abedin et al., 2015), while the wide part of the stream is connected with comet 96P/Machholz (Hasegawa, 1979, Abedin et al., 2018). The age and formation mechanism of the Quadrantid meteoroid stream core and the relationship with the asteroid (196256) 2003 EH1 have been studied previously by several authors (Jenniskens, 2004; Williams et al., 2004a; Wiegert & Brown, 2005; Abedin et al., 2015).

Here we do not analyse it on a million-year timespan, because most authors (Jenniskens, 2004; Wiegert&Brown, 2005; Kaňuchová and Neslušan, 2007; Abedin et al., 2015) have noted the similarity between the current orbit of asteroid (196256) 2003 EH1 and the mean orbit of Quadrantids and identified asteroid as the parent of the meteoroid shower. They have concluded that the most likely time period for the formation of the central portion of the stream is circa 1750–1800 AD.

The evolution of the Quadrantid stream orbit and its individual meteoroids has been studied repeatedly, and the first work in this series was Hamid & Youssef (1963). Hughes et al. (1981) showed that the nodes motion is very sensitive to the orbital parameters used, and in works (Gonczi et al., 1992; Froeschlé, 1986; Williams & Wu, 1993) it was shown that the evolving orbit experiences multiple close encounters with Jupiter and behaves randomly. From this it follows that meteoroids, which in the phase space of the orbits were initially close, can move away from each other at a considerable distance in a short period of time.

The orbital evolution of the asteroid (196256) 2003 EH1 has been studied by many authors, most recently by Kaˇnuchová and Neslušan (2007), Kholshevnikov et al. (2016), Galushina et al. (2017). Numerical integration of motion equations of the asteroid 2003EH1 and its 500 clones (Williams et al., 2004b; Galushina et al., 2017) showed that the orbits of the clones due to multiple close encounters with the planets, move away from the nominal orbit very quickly, which confirms the above. Consequently, in case of Quadrantid, any initial structure will be lost very quickly.

However, we will illustrate this fact clearly. To do this, we create a model that reflects the actual dynamical properties of the stream, but does not reflect ejection process. We simulated the emission of Quadrantid test particles in the perihelion of the 2003EH1 asteroid orbit with the same velocities calculated using the Whipple formula (Whipple, 1951). Ejection was simulated isotropic. This is described in detail in section 2.

The questions we will answer in this work are: (1) Are the orbits of the meteoroid stream stable for long? (2) How is the period in which the orbits can be regarded as stable long? (3) What is the cause of chaos in the motion of meteoroid stream? (4) What resonances are manifested in the Quadrantid meteoroid stream? The simple dynamic approach we use does not entail discussion of physical mechanisms. Two things only are important for our modeling: the location of the outbursts (i.e. perihelion), and the exact timing of the ejected particles.

**2. Modeling meteoroid ejection.**

*2.1 Model*

Stream modeling method is well understood due to cometary volatiles sublimation, and it was reported many times by many authors (Whipple, 1951; Jones, 1995; Crifo and Rodionov, 1997; Hughes, 2000; Brown & Arlt 2000; Asher & Emel'yanenko, 2002; Ryabova, 2016). All meteoroid ejection models share the same physical concepts to a great extent, though with slight modifications. The main idea of this method is that we generate an ejection of some amount of test particles at some points on the orbit of the parent body and follow the evolution of each of them to the epoch of our choice. Current dust production from (196256)2003 EH1 is too small (Kasuga & Jewitt, 2015) to supply the mass of the Quadrantids during interval of 200-500 years ago. If 2003 EH1 is the source of the Quadrantid core, then mass must have been delivered episodically.

Existing dust ejection models from comet nuclei suggest that the mass loss is made up of both gas and dust, the dust being given a small velocity (1-100m/s) with respect to the cometary nucleus. The ejection velocities are much smaller (few tens hundreds m/s) than the orbital speed of the comet (few tens of km/s at perihelion). Abedin et.al. give the lowest relative velocities within the clones of the core Quadrantids and 2003 EH1, range from 200 m/s to 800 m/s, with the majority of the clones having relative velocities exceeding 1 km/s as the characteristic particle speed. We seem a reasonable compromise the average value produce speeds in the 200–800 m/s range.

We based on the above-mentioned results and chose six "ejection epochs". We simulate ejections of model meteoroids at perihelion of the parent orbit approximately around 1760 – 1790 AD. In this experiment we confined ourselves to the following points:

1. The timing of the appearance in the sky (around 1835 AD) (Quetelet,1839; Fisher,1930).

2. We assumed that the asteroid 2003EH1 is the "core" of the stream (Abedin et.al., 2015; Kholshevnikov et.al., 2016).

3. The nominal orbit of the asteroid 2003EH1 and orbits of the confidence ellipsoid can be considered regular on the time interval 1760-2003.

4. The results of many authors (Williams et al., 2004b; Wiegert & Brown, 2005; Micheli et al., 2008, Sambarov et.al., 2018) excluded the proposed identification of comets C/1490 Y1 (Ki-Won, 2009) and C/1385 U1 (Ho, 1962) as the historical cometary apparitions of asteroid 2003 EH1; this excludes that the meteor shower could be the result of two large decays of comets around 1400 and 1500.

The numerical integration of test-particle orbits of the meteoroid stream studied here were performed with the Everhart (1974) 19th-order procedure with variable step size from the moment of ejection to the present day, in a model Solar system which was described in detail in the works (Galushina et al., 2015, Galushina & Sambarov, 2017, Sambarov & Syusina, 2018, Galushina & Sambarov, 2019). Planetary positions were taken from the Jet Propulsion Laboratory (JPL) Planetary development Ephemeris – DE431.

The length of the backward integration of 1750 years is chosen on the basis of the assumption of relatively young age of the central portion of the Quadrantid meteoroid stream (Jenniskens, 1997; Wiegert & Brown, 2005, Abedin et.al., 2015) and because we found (Galushina & Sambarov, 2017, Sambarov et.al, 2018) that, according to the researches of chaoticity and dynamical evolution of asteroid (196256) 2003 EH1 with 500 clones, the stream orbits do not significantly disperse until 1700 AD. It shows that the orbit may be considered as regular on this time interval.

## 2.2. Evolution of the model meteoroid stream

### 2.2.1. The short-term orbital evolution

This section presents the short-term variability of the orbital parameters of the modeling Quadrantid meteor stream by investigating the motion of 100 test particles spaced around the parent body at different epochs in a single outburst. We simulate ejections of 600 model meteoroids at perihelion passage of the parent body approximately 1764, 1769, 1775, 1780, 1785, 1791 AD. We investigated the orbital evolution of each ejected particle from the moment of ejection to the present day. The probabilistic orbital evolution study results are presented in **Fig.1**. The **Fig.1** shows evolution of semi-major axis *a*, eccentricity *e*, inclination *i*, longitude of ascending node $\Omega$, and argument of perihelion $\omega$. The evolution of model particles is shown in gray, and the asteroid orbit is marked out in black. The semi-major axis remains at a relatively stable value close to 3.1 au. This suggests that the orbital energy remains constant throughout the interval despite a number of close encounters with planets. The behavior of model particles and (196256) 2003 EH1 are not characterized by large variations in the eccentricity and the inclination on the period 1780-2019 yr.

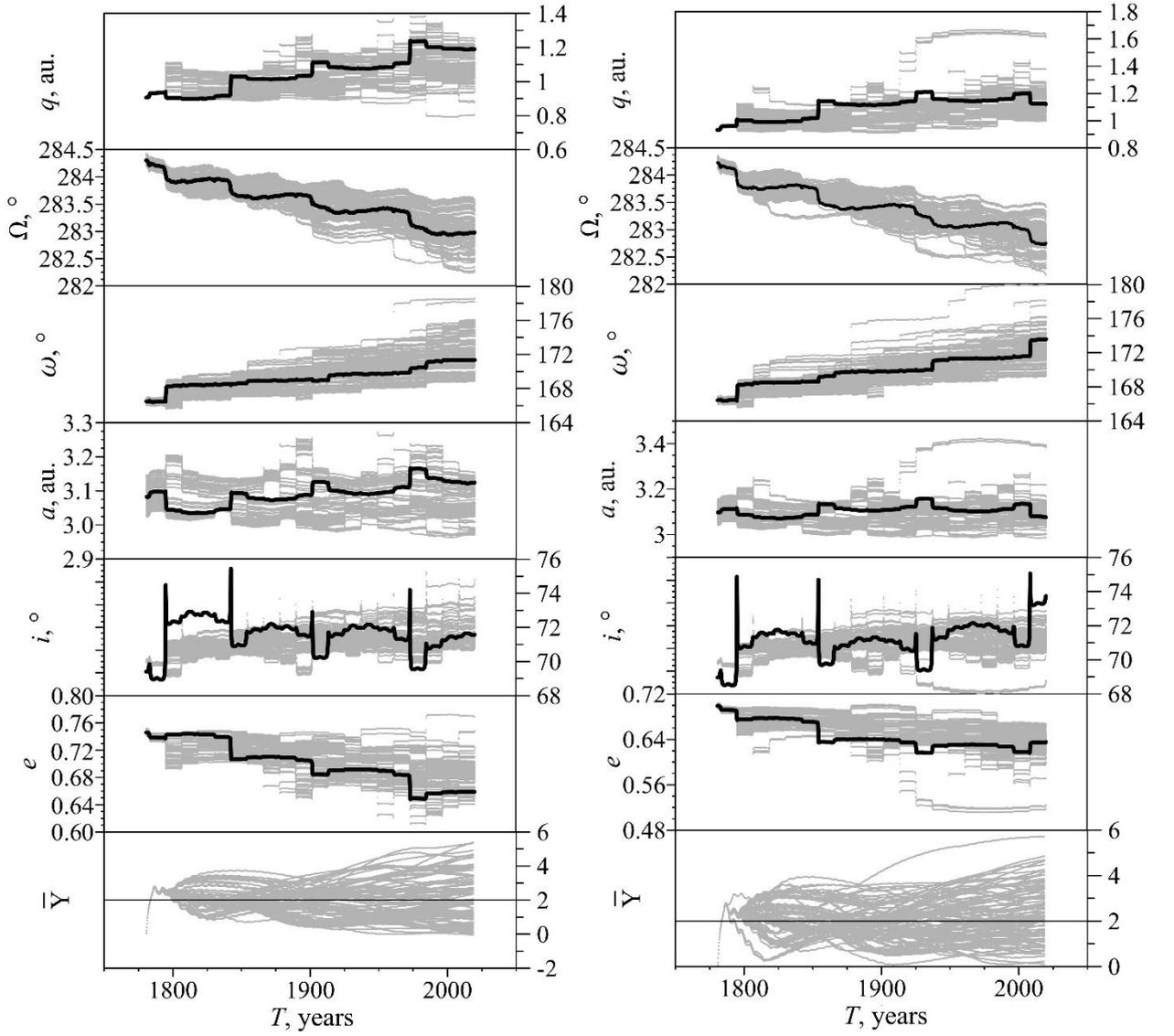

**Fig 1.** Changes in the six orbital elements $q$, $\Omega$, $\omega$, $a$, $i$, $e$ and evolution of the parameter MEGNO of the model Quadrantid stream (100 gray particles) and asteroid (196256) 2003 EH1 (black) between the moment of ejection around 1780 (left) and around 1786 (right) date and 2019 AD.

This suggests that the orbital energy has remained constant throughout the interval despite a number of close encounters with planets, as mentioned above. It is very interesting that the semi-major axis of the orbit is approximately conserved, while the orbital eccentricity and inclination oscillate out of phase. The perihelion distance $q$ shows fairly smooth monotonic change on entire interval. It was significantly smaller in the past, having a value of $q \sim 0.9$ in the moment ejection. Most of the test particles have slightly smaller values than those of the asteroid orbit.

The variations in the orbital parameters of model meteoroids remain within the small range for interval prior.

The MEGNO parameter $\bar{Y}(t)$ allows the chaotic and regular motions to be separated and their predictability time to be estimated (Cincotta et al., 2003). The time evolution of $\bar{Y}(t)$ value manifests certain features specific to different types of orbits. The MEGNO parameter grows linearly with time for a chaotic orbit, whereas for the regular orbit, this parameter is less than two (or oscillates about this value). For example, it is known that for quasi-periodic (regular) orbits $\bar{Y}(t)$ always tends towards 2. $\bar{Y}(t)$ for stable orbits of a harmonic oscillator type equals zero.

As is seen in **Fig.1**, the parameter $\bar{Y}(t)$ linearly grows for half of the simulated particles, passes through value 2, and then the motion becomes unpredictable for slightly less than half of model meteoroids ejected after 1780 AD from (196256) 2003 EH1. These particles move in vicinity of the following mean motion resonance 2:1J with Jupiter. The Quadrantid stream has long been known to be located near this resonance (Hughes et. al., 1977). The orbital elements (nominal semi-major axis) show a periodicity with a roughly 59 yr period. Similar results were obtained by the authors (Hughes et al., 1981), which been observed in simulations of the Quadrantid stream. This periodicity associated with the 2:1J resonance has been analytically determined by Murray (Murray & Dermott, §8.9, 1999; Murray, 1982). The test particles ejected from (196256) 2003 EH1, in view of the resonance may have further effects on the Quadrantid stream, because they could find themselves within this resonance.

In extended simulations, some ejected particles are seen to have been in the 2:1J resonance in the more distant future (**Fig. 2**). Non-resonant states the proximity of the mean motion resonances 2:1J have a strong influence on the motion of model meteoroids. When the evolution of the nominal orbit and a stable particle orbit is considered, the elements are found to undergo a series of periodic, low-amplitude changes. Another situation is observed for unstable particle. Orbit elements, especially semi-major axis, suffer fast large changes as a result of very close approaches to Jupiter. This effect is illustrated for the same particles (**Fig.3**). For example, in **Fig.3** (the right-hand side set of panels), unstable ejected particle has *a* after date 1780 AD differing by about 0.2 au from a before date 2600, because Jupiter has caused *a* to oscillate significantly while the particle is in the 2:1J resonance. The critical argument sometimes circulates, sometimes librates (1780-2020 yr.) with large amplitude. So, we can say that the particle is in vicinity of the resonances but is not captured by them. Mean motion resonance is unstable and may be reason of the chaotic motion (Chirikov, 1979, Bordovitsyna et. al, 2012, Bordovitsyna et. al, 2014). The resonance argument (**Fig.3**) of the asteroid (196256) 2003 EH1 puts it near, but not currently in, the 2:1J mean motion resonance with Jupiter. **Fig. 3** shows that neither the nominal orbit of (196256) 2003 EH1 no its ejected particles are in the strict 2:1J mean-motion resonance during the simulations shown.

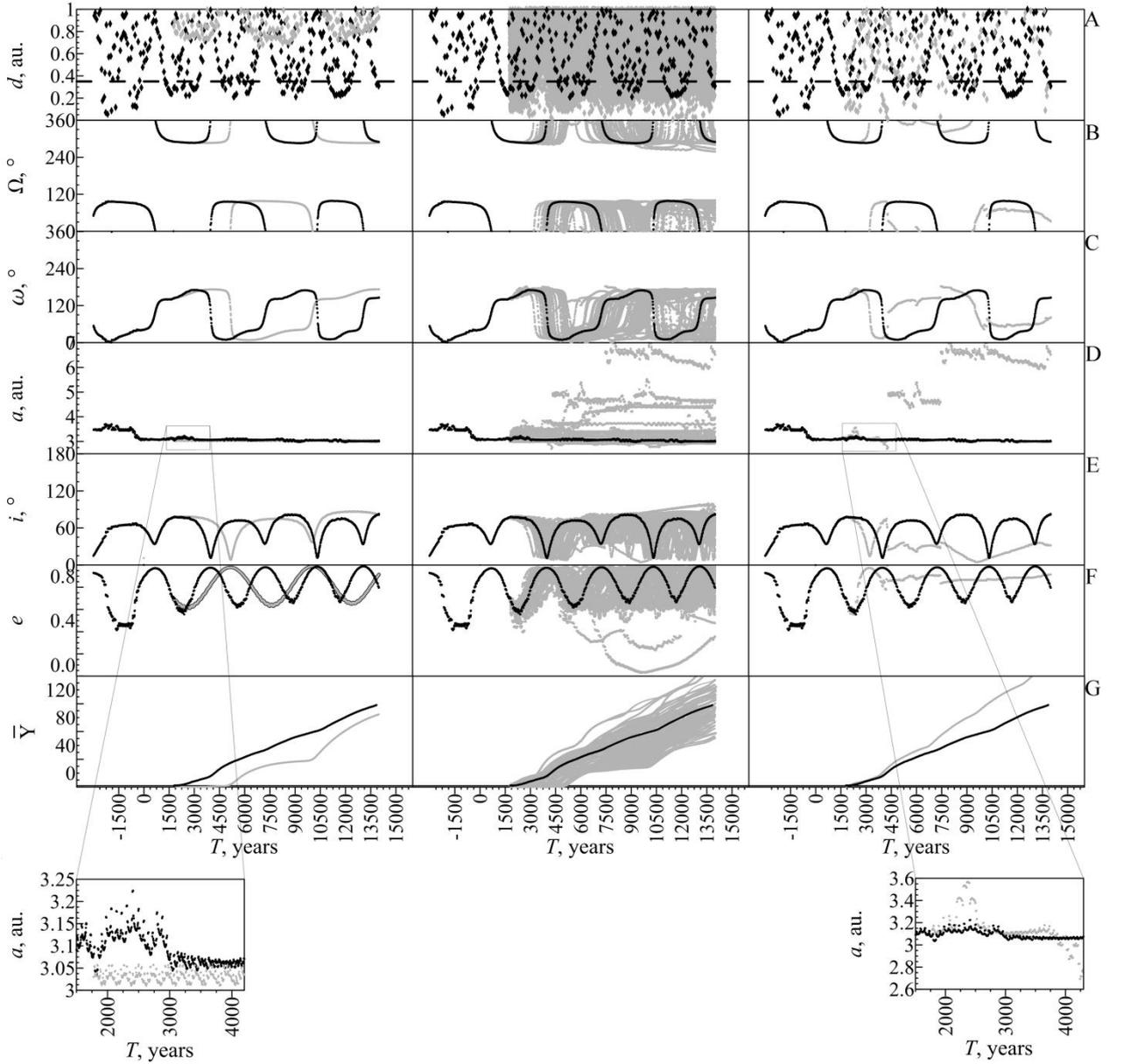

**Fig 2.** Evolution of the values of relevant parameters for the nominal orbit of asteroid (196256) 2003EH1 (black dots) and the model Quadrantid stream (center set of panel is a version of 100 grey particles; the left-hand side set of panels is a version of stable ejected particle; the right-hand side set of panels is a version of unstable ejected particle) distance from Jupiter (top panel, with Hill radius of Jupiter, 0.35 au), changes in the six orbital elements Ω, ω, $a$, $i$, $e$ and evolution of the parameter MEGNO (bottom panel).

The same scenario for the MEGNO parameter $\bar{Y}(t)$ can be seen for meteoroids ejected after the moment ejection from (196256) 2003 EH1 (**Fig.4** and **Fig.5**). The motion also becomes unpredictable for slightly less than half the simulated particles. For the remaining particles, which have a parameter $\bar{Y}(t)$ less than 2, the objects motion remains quasi-periodic. The authors (Abedin et al., 2015) note that integration for 200-300 years back in time is beyond the Lyapunov time. They concluded that the results from forward integration's should be treated with caution, in particular when the initial orbital elements of the parent are selected from backward integration beyond the Lyapunov time. The author (Tancredi, 1998) showed that the motion of NEAs and Jupiter family comets have been considered to be chaotic in a short time scale, typically 50-100 years, and the close encounters with Jupiter are supposed to be the cause of the fast chaos. On the basis of this information it is easy to conclude that we do not expect to be able to compute the position of bodies within their orbits beyond a few times in this interval. But attention should be drawn to the fact that "weak chaos" is largely confined to the true anomaly. Thus, the shape of the orbit can be computed reliably over much longer time scales than can the body's position within the orbit.

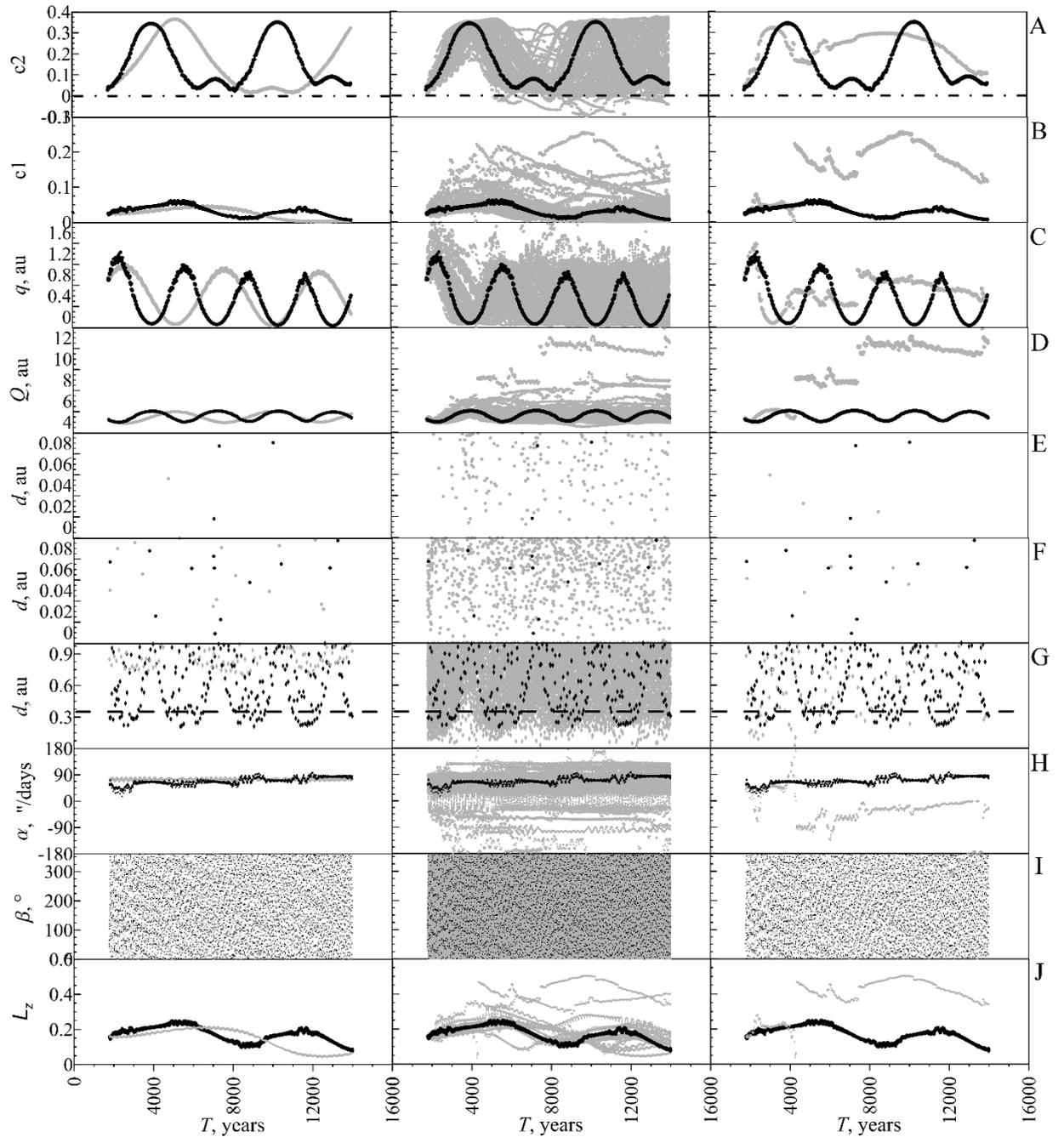

**Fig 3.** Evolution of the values of relevant parameters for the nominal orbit of asteroid (196256) 2003EH1 (black dots and lines) and the model Quadrantid stream (center set of panel is a version of 100 grey particles; the left-hand side set of panels is a version of stable ejected particle; the right-hand side set of panels is a version of unstable ejected particle ) condition c2 (A-panel) and c1 (B-panel), perihelion $q$ (top panels, C-panels) and aphelion $Q$ (D-panels) distances, distance from the Earth (E-panels), distance from Jupiter (F-panels, with Hill radius of Jupiter, 0.35 au), the evolution of the resonant band α (H-panels), the critical argument β (I-panels), value of the Lidov–Kozai parameter (bottom panels, J-panels)

*2.2.2. The long-term orbital evolution*

It is to be noted that the evolution of simulation particles of meteor stream does not show the Lidov-Kozai mechanism on a short period, as asteroid (196256) 2003EH1 and some simulation particles for the long-period time (**Fig. 2**). Although for this mechanism to work the aphelion of a massive perturber located relatively close to the aphelion distance of the affected object and this putative perturber should probably move in a low-eccentricity, low inclination orbit. The reason for that may be due to close encounters with Jupiter or proximity of the orbit of the asteroid to mean motion resonance nominal location. For a short period of time the secular resonances were not able to manifest for ejection particles. However, certain effects of the secular resonances might become apparent only over a longer period of time. The dynamical evolution of the Quadrantids has been studied by several authors in the past. Some of them (e.g. Williams et. al. 1979; McIntosh, 1990; Neslusan et al. 2013; Abedin, 2018) found a libration of perihelion distance of the orbits of particles, representing the Quadrantid meteoroids, with the period of about 4000 years. Also, argument of perihelion and longitude of ascending node vary. Because of the significantly longer period of the libration cycle than the explanation provided above integration time, one cannot be sure whether the increase of the mean value of the MEGNO parameter, $\overline{Y}$, (the last panels in **Fig.1, Fig.4** and **Fig.5**) of some test particles is permanent or only temporary and, hence showing, whether the orbits are chaotic or regular.

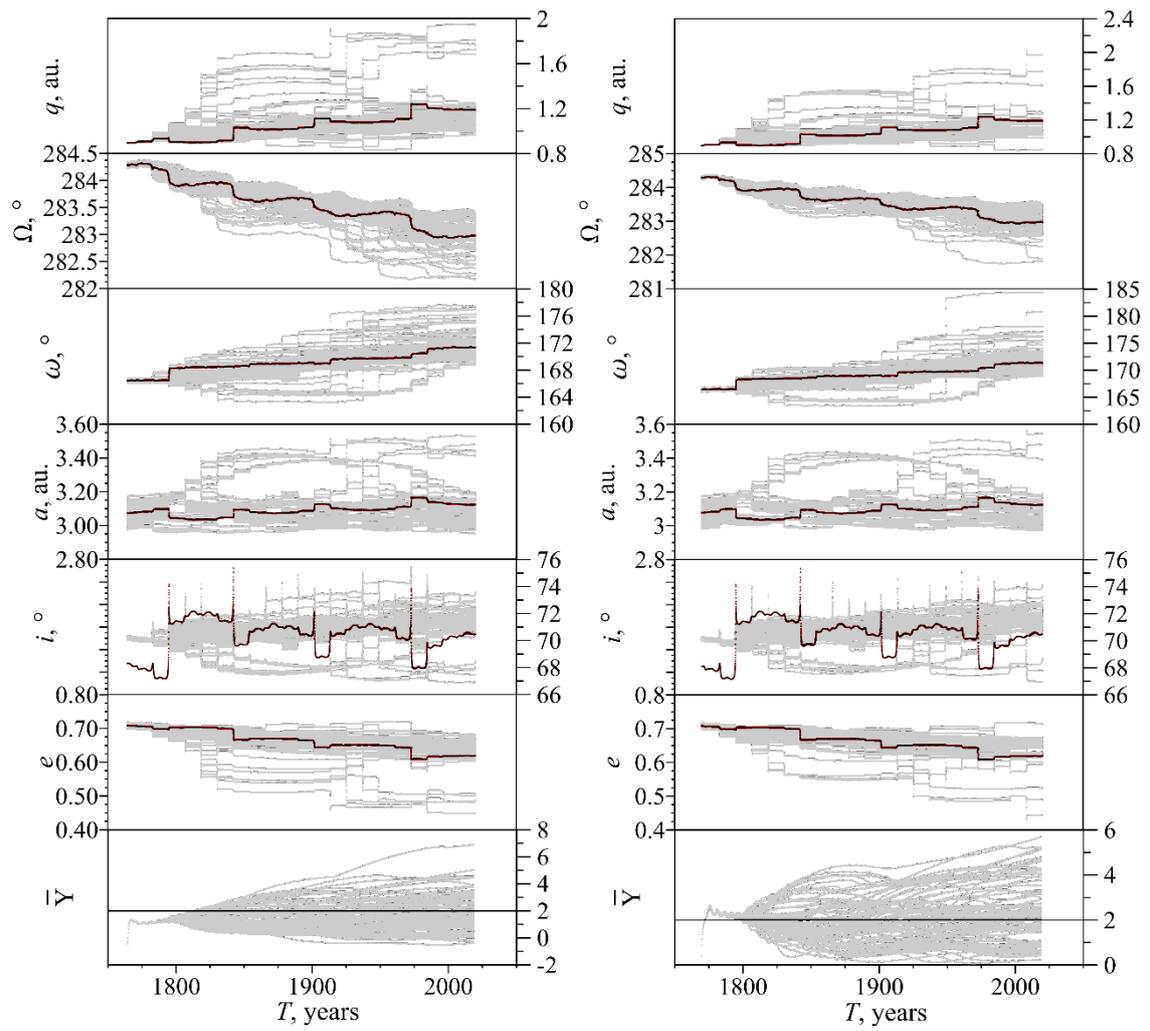

**Fig 4.** Same as Fig. 1 but for the moment of ejection around 1764 (left) and around 1769 (right).

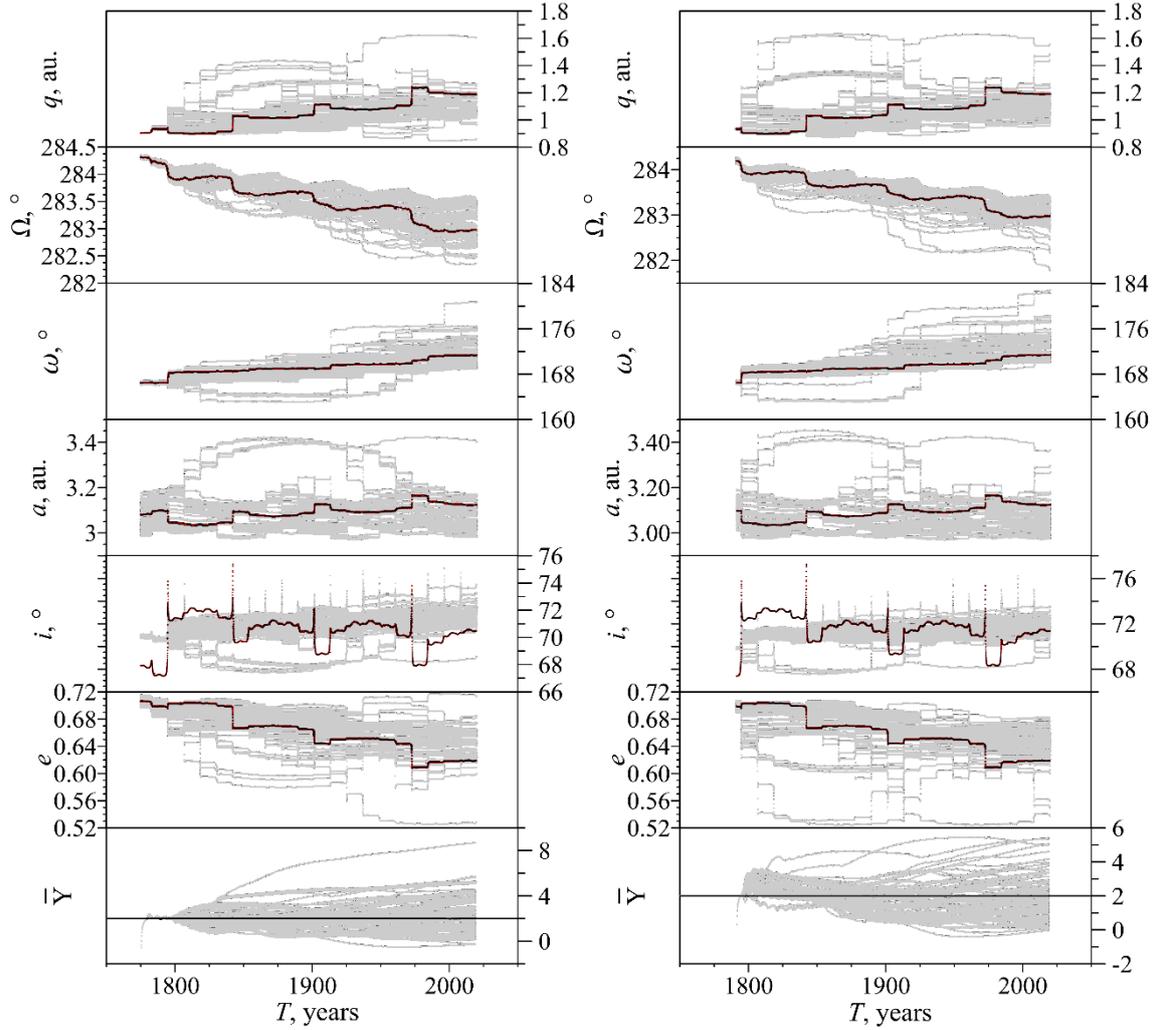

**Fig 5.** Same as Fig. 1 but for the moment of ejection around 1775 (left) and around 1791 (right).

The character of the peculiar dynamics of unstable simulated particle shows several passages through the 2:1J mean-motion resonance in time intervals between 1900-2100 years and 8000-12000 years (**Fig. 3H**, right panel). On this interval, the large amplitude libration of the resonant band value $\alpha = 2n_a - n_5$ relative to 0 takes place, where $n_a$ is a mean motion of the asteroid, $n_5$ is a mean motion of Jupiter. One more amazing feature is that, before the year 4000, the libration center of the resonance characteristic α with a small amplitude of oscillations was shifted to the positive side (80″ per day), while it moved to the negative side (−80″ per day) after the year 4000. In other words, for the entire interval under consideration, unstable simulated particle is moving in the resonance vicinity, but passes through the sharp resonance only a few times. The unstable geometric configuration of the the 2:1J mean-motion resonance with Jupiter for unstable simulated particle (an unsuccessful attempt to capture the object into resonance), more precisely, the transition between the resonant and nonresonant states, may be a cause of such a strong manifestation of the chaotic character of the object's motion, as that shown in **Fig. 1, Fig. 3, Fig. 4, Fig. 5** (first to bottom panel), where the MEGNO parameter passes a value of 2 starts to rise steeply. The libration center of the resonance characteristic α with small amplitude of oscillations for stable simulated particle is always on the one the positive side (**Fig. 3H**, left panel).

Unfortunately, the integration period of less than 300 years is too a short to decide whether the considered set of orbits is in a regular regime or the orbits are chaotic and their motion is unpredictable for a longer time. An application of the MEGNO to evaluate a regular or chaotic behavior of a set of orbits requires to follow the dynamics of the system during a sufficiently long period. In the longer-term the behavior of the Quadrantid stream characterized by large variations in $e$ and $i$. These have been seen by a number of investigators, starting with Hamid and Youssef (1963). The large oscillations are similar to those associated with the Lidov-Kozai mechanism (Kozai, 1962; Lidov, 1962). In fact, the evolution of stable ejected particles exhibit Kozai-type circulation in that their swings in $e$ and $i$ approximately conserve $a$ (**Fig.2**) and $L_z = \sqrt{1-e^2}\cos i$ (**Fig. 3**). Top left-hand side panel in **Fig. 2** show that the peculiar dynamics of stable simulated particle leads to avoiding close encounters within the Hill radius of Jupiter for the most part. The value of the semi-major axis (fourth to top panel) changes slightly during the time interval displayed, drifting from one quasi-constant value to a relatively close one (also quasiconstant) for most of the integrated time. On the other hand, the values of $e$ (fourth to bottom panel) and $i$ (third to bottom panel) oscillate, alternating high $e$ and $i$. The libration in ω confirms that this object is currently subjected to a Lidov–Kozai mechanism. But after the transition of the argument of periapsis ω from libration around 180◦ (in the range of 500 years) to libration around 0◦ and back the parameter MEGNO begins to grow slowly. When particles are far away from the Jupiter, the orbital evolution changes only slightly and becomes smooth (**Fig.**

**3**) as the kinks disappear. Such behavior is often found among near-Earth asteroids (NEAs) with values of their semi-major axes close to that of the Earth that move confined between the orbits of Venus and Mars (Michel & Thomas 1996; de la Fuente Marcos & de la Fuente Marcos 2015), but in this case the value of ω also oscillates. Therefore, what is found is consistent with a Lidov-Kozai mechanism state. In our case, the Lidov–Kozai mechanism is at work to avoiding close encounters with Jupiter at aphelion for stable ejected particles. Top left-hand side panel in **Fig. 2** show that the peculiar dynamics of stable simulated particle leads to avoiding close encounters within the Hill radius of Jupiter for the most part. This simulated particle does not have many frequent close encounters with Jupiter and displaying clear episodes of Lidov–Kozai mechanism behavior.

**Fig. 2,** left-hand side set of panels, shows a libration of the argument of periapsis of the simulation particles orbits, representing the Quadrantid meteoroids, with the period of about 4000 years. Also, eccentricity and inclination node vary, exchanging energy with each other, for some particles. We may see at longer period that the increase of the mean value of the MEGNO parameter, $\overline{Y}$, of test particle is permanent and, hence, the orbit is stable. **Fig. 2**, right-hand side set of panels, shows the long-term evolution for unstable simulated particle has many frequent close encounters with Jupiter in the area of the Hill sphere radius of Jupiter, 0.35 au, is displayed by grey horizontal line. The dynamical evolution of the simulated particle is far more chaotic. The values of the various parameters change more rapidly and by a wider margin. Close encounters with Jupiter systematically enable about 1800-2000 years into the past (but also prior to 3800-4200 years into the future) and the value of the Lidov–Kozai parameter changes within a larger interval as well (**Fig.3**). Similarly, close encounters with Jupiter are not treated by the Kozai formalism and can transfer particles away from the Kozai trajectories over time. In our case, ejected particles might be controlled by direct perturber Jupiter.

The author (Naoz, 2016) showed that the Lidov–Kozai mechanism can excite the value of the eccentricity for a long period of time until collisions with inner or outer perturbing bodies are possible. Although the orbital evolution of some ejected particles appears to contribute to enhance the stability of the orbits in the sense of keeping them within a well-defined section of the available orbital parameter space, **Fig. 2** indicates that the eccentricity may eventually reach values that lead the aphelion closer to Jupiter. Conversely, a larger value of the eccentricity may lead to a shorter perihelion and an eventual collision with the Sun or an ejection from the Solar system or into its outskirts. In our case, the value of the aphelion distance, that oscillates close Jupiter (see **Fig. 3**), opening the door to eventual impacts our ejected particles if one of the nodes reaches aphelion (Naoz, 2016; Ribeiro et al. 2016). Unstable ejected particles can eventually become impactors as a result of the complex orbital evolution discussed here and in the event of an impact, they will most probably come from out of Sun pertubations or solar radiation forces. Within the Lidov-Kozai mechanism, many ejected particles would be protected against close encounters with Jupiter and the Sun. Here, we show that the orbits of some ejected particles are strongly affected by the Lidov-Kozai mechanism.

Among mentioned librational ejected particles' orbits, the conditions for the Lidov–Kozai resonance (Vashkov'yak and Teslenko, 2008) $c1 = (1 - e^2)Cos^2 i \leq 3/5$, $c2 = e^2(2/5 - Sin^2 i\ Sin^2 \omega) < 0$, are satisfied only for some orbits (**Fig. 3**). The elements of the remaining ejected particles mentioned orbits satisfy the condition $0 \approx c2 > 0$, where the actual libration $\omega$ is explained from the fact that, for majority orbits, c1 is close to its bifurcational value 3/5, while c2 (being positive) is found to be close to zero (**Fig. 3**). Under these conditions, during the process of the evolution of the satellite orbits, it is possible that a significant change in the eccentricity runs in the antiphase change of sin $i$.

3. **Discussion and conclusions**

In this paper, we have presented one of the possible scenarios for the dynamical evolution of the meteoroid stream formed by the asteroid (196256) 2003 EH1. Our analysis of the dynamics of meteoroid particles assumed to be released from the near-Earth asteroid (196256) 2003 EH1 revealed the complicated dynamical structure of its meteoroid stream, which approaches Jupiter's and Earth's orbits. Meteoroids inherit the dynamic properties of the asteroid but not all. The slight timing inconsistency is perhaps due to small number statistics. We suppose that the ejected particles' orbits can be captured in resonances as several known asteroids or comets. In fact, the meteoroid stream can be regarded as a unique structure originating, most likely, in cometary and asteroid parent bodies. There are several tens of Jupiter-family comets, many of which have dynamical properties similar to those of (196256) 2003 EH1, and several known asteroids with their perihelion in the interior of Earth's orbit. Discovering and studying them is a challenge for future meteor research.

We have studied the long-term dynamical evolution of orbital solutions similar to that of present day ejected particles from asteroid (196256) 2003 EH1. On average, slightly less than half of the ejected particles have chaotic motion from simulations of the orbit of this body and of meteoroids released from it at different intervals in the past. We suppose that the reasons are the frequent close approaches of the ejected particles with Jupiter and they located near mean motion resonance 2:1J with Jupiter. The motion of these objects has considered to be chaotic in a long-time scale, and the close encounters with Jupiter are supposed to be the cause of the faster chaos. Another reason is that a non-resonant state near the mean motion resonance 2:1J has a strong influence on the motion of the Quadrantid meteor stream. This "weak chaos" is largely confined to the true anomaly. Consequently, the shape of the orbit can be computed reliably over much longer time scales than can the body's position within the orbit.

The parameter MEGNO increases for the simulation particles of meteor stream moving around the Sun and perturbed by the planets. High value of the parameter MEGNO are due to frequent changes in semi-major axis induced by multiple close encounters with Jupiter near Hill sphere. We finally note that the chaotic behavior of the simulation particles of meteor stream may be caused not only by close encounter with planets but also by unstable mean motion or secular resonances.

Within the Kozai-Lidov mechanism, many ejected particles would be protected against close encounters with Jupiter. Here, we show that the orbits of some ejected particles are strongly affected by the Kozai-Lidov resonance. Our calculations show that stable ejected particles can be trapped in a Lidov–Kozai resonance that protects them from close encounters with Jupiter. Lack of close encounters with Jupiter leads to a rather smooth path with nearly constant semi-major axis. The coupled oscillation of the three orbital parameters, $e$, $i$, and $\omega$, for stable ejected particles is observed. The value of the argument of perihelion mostly librates about $90°$. However, close encounters with Jupiter are not treated by the Kozai formalism and can transfer particles away from the Kozai trajectories over time, as can be seen from the figure for unstable ejected particles.

What can cause activity in the inactive parent of a meteor shower? The orbit itself is stable in the century preceding our proposed meteor emission points. Thus, particles ejected from the asteroid 2003EH1 ~250 years ago produces a stream with orbital characteristics matching those of the Quadrantid meteoroid stream. The first close approaches between many of the ejected particles and the Earth, argues in favor of formation in either a cometary disintegration or the result of transient activity of the comet over a short period of time. In addition, over the past few hundred years, the perihelion distance of the parent asteroid (196256) 2003 EH1 was relatively large, making impossible the solar heating to be large enough to cause a cometary activity.

For future work, it would be instructive to do a detailed abstract study, with high-precision Quadrantid orbits and integrated their orbits backward in time, along with the analysis of the dynamics structure, we could identify the most likely age of the core of the Quadrantid meteoroid stream.

**Acknowledgements**

One of the authors (SGE) would like to acknowledge Ryabova G.O. and L. Kornos for interesting remarks at the conference IMC-2018 (International Meteor Conference, August 30–September 2, Pezinok-Modra, Slovakia) and thank for very interesting discussions with A. Sekhar and D. J. Asher at the conference Meteoroids-2019. This research has made use of data and/or services provided by the International Astronomical Union's Minor Planet Center. This research has made use of NASA's Astrophysics Data System and Scopus (Elsevier).